\documentclass[twocolumn]{aastex62}

\graphicspath{{./}{figs/}}

\usepackage{comment,psfig}

\shorttitle{The Origin of Radio Emission from Radio-Quiet AGN}
\shortauthors{F. Panessa, R.~D. Baldi, A. Laor, P. Padovani, E. Behar, I. Mchardy}

\begin{document}

\title{The Origin of Radio Emission from Radio-Quiet AGN}

\author{F. Panessa} 
\affiliation{INAF – Istituto di Astrofisica e Planetologia Spaziali, via Fosso del Cavaliere 100, I-00133 Roma, Italy}
\author{R.~D. Baldi}
\affiliation{School of Physics and Astronomy, University of Southampton, Southampton, SO17 1BJ, UK}
\affiliation{Physics Department, Technion, Haifa 32000, Israel}
\author{A. Laor} 
\affiliation{Physics Department, Technion, Haifa 32000, Israel}
\author{P. Padovani}
\affiliation{European Southern Observatory, Karl-Schwarzschild-Str. 2, D-85748 Garching bei M\"{u}nchen, Germany}
\author{E. Behar} 
\affiliation{Physics Department, Technion, Haifa 32000, Israel}
\author{I. Mchardy}
\affiliation{School of Physics and Astronomy, University of Southampton, Southampton, SO17 1BJ, UK}




\begin{abstract}

The central nuclei of galaxies, where super-massive black holes (SMBHs) are thought to reside, can experience phases of activity when they become Active Galactic Nuclei (AGN). An AGN can eject winds, jets, and produce radiation across the entire electromagnetic spectrum. The fraction of the bolometric emission in the radio spans a factor of $\sim$10$^{5}$ across the different AGN classes. The weakest radio sources, radio-quiet (RQ) AGN, are typically 1,000 times fainter than the radio-loud (RL) AGN, and represent the majority of the AGN population. In RL AGN, radio emission is essentially all produced by synchrotron emission from a relativistic jet. In contrast, in RQ AGN the absence of luminous jets allows us to probe radio emission from a wide range of possible mechanisms, from the host galaxy kpc scale down to the innermost region near the SMBHs: star formation, AGN driven wind, free-free emission from photo-ionized gas, low power jet, and the innermost accretion disc coronal activity. All these mechanisms can now be probed with unprecedented precision and spatial resolution, thanks to the current and forthcoming generation of highly sensitive radio arrays.

\end{abstract}



\
\section{Introduction} \label{sec:intro}

The evolution of each massive galaxy in the Universe is strictly connected to the activity of the super-massive black hole (SMBH, 10$^{6-10}$ M$_{\odot}$) located at its centre. The SMBH becomes active when surrounding material is captured by the BH gravitational potential well forming an accretion disc and occasionally ejecting plasma in the form of outflow and/or a collimated jet. In this phase, active nuclei emit light over the entire electromagnetic spectrum due to a large variety of physical mechanisms.

Only $\sim$ 10\% of Active Galactic Nuclei (AGN) possess the ability of launching powerful  relativistic jets that shine in the radio band and emit synchrotron radiation (e.g., \citealt{begelman84}). These have been called Radio-Loud (RL) AGN, as the radio emission is typically 10$^{3}$ times brighter than that in Radio-Quiet (RQ) AGN$\footnote{The classical radio-optical radio-loudness parameter R is defined as {\it f}(4400 \AA)/{\it f}(6 cm) \citep{kellermann89}, where RQ AGN have R $<$ 10. See also \citet{terashima03} for a radio--X-ray based definition R$_{X}$ = $\nu$L$_{\nu}$(6 cm)/L(2-10 keV).}$. However, a sharp separation between RL and RQ is hard to set, as AGN show a large variety of radio properties and morphologies, with sources exhibiting compact cores, jets, knots, and extended diffuse emission in a wide range of strengths and sizes ranging from sub-pc up to kpc and even Mpc scales. The radio emission from these components depends on frequency and on the AGN type.

\begin{figure*}
\centering
  \includegraphics[width=1.0\linewidth]{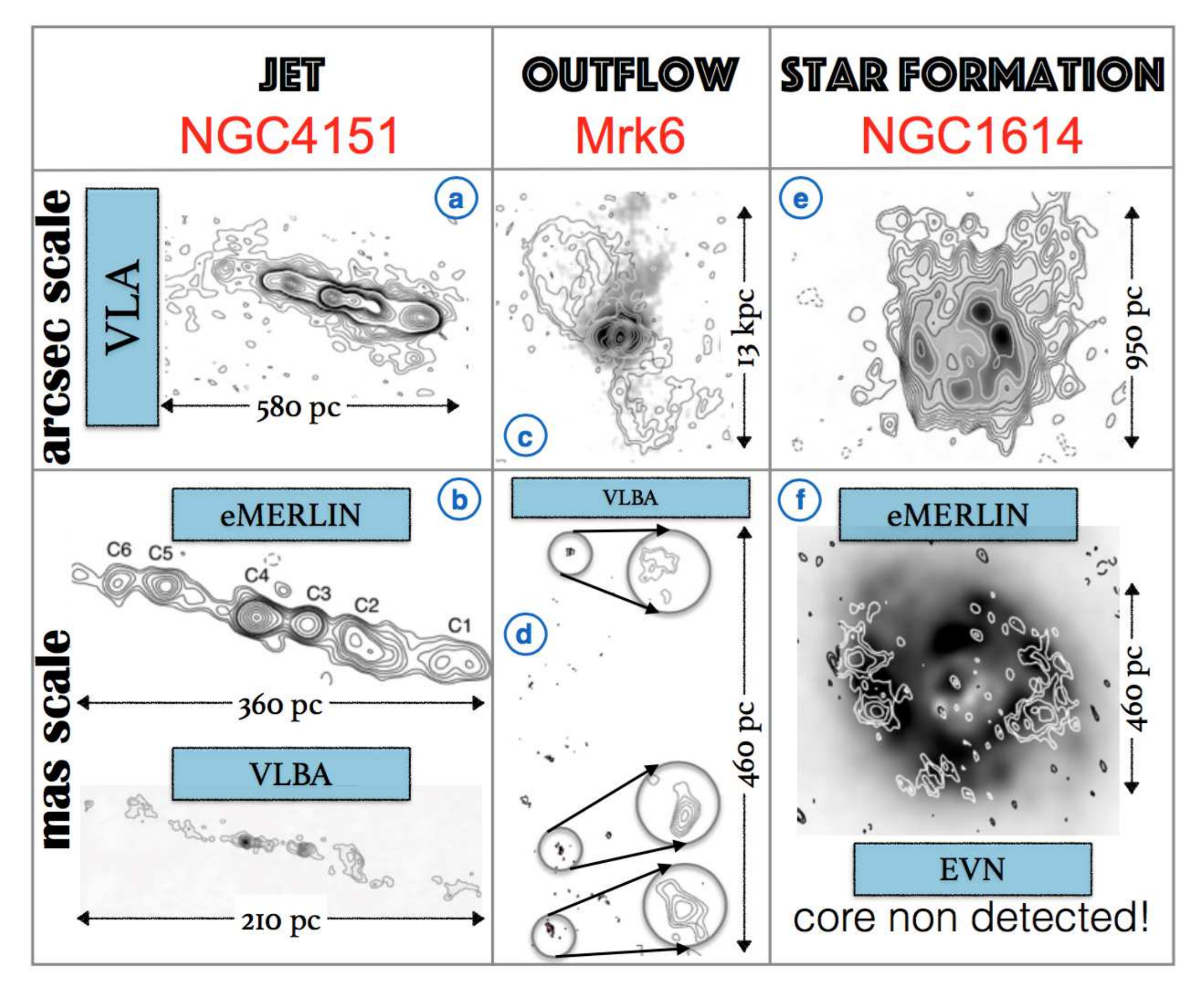}
  \caption{Prototype radio maps of RQ AGN with extended radio morphologies. Upper panel maps with arcsec resolution (from VLA) and lower panel maps with mas resolution: 8-GHz jetted structure of NGC~4151 (panel $a$, \citealt{pedlar93}), resolved in several components (panel $b$) with eMERLIN \citep{williams17} and VLBA \citep{mundell03} at 1.4 GHz; Mrk~6 at 1.5 GHz (panel $c$, \citealt{kharb06}) for an outflowing bubble-like structure resolved in smaller components with VLBA at 1.6 GHz (panel $d$, \citealt{kharb14}); NGC~1614 at 8.4 GHz (panel $e$, \citealt{olsson10}) from a star-forming diffuse structure, showing a clear ring with eMERLIN at 5 GHz (panel $f$, \citealt{olsson10}), no core detected with EVN at 5 GHz \citep{herrero17}.}
  \label{fig:examples}
\end{figure*}

Radio emission, especially at high frequencies, is highly penetrative (at variance with the optical and soft X-ray bands, which are heavily absorbed in at least half of the AGN) fostering an unbiased view of the AGN population. The radio power, morphology and spectral index provide an initial characterization of the emission, which largely depends on the spatial resolution of the used radio interferometer (arcseconds to milli-arcseconds, see Figure \ref{fig:examples}) and on the redshift of the source (local to high z). At low redshift, interferometric studies allow us to investigate the radio nuclear properties and to probe the plethora of physical mechanisms acting in RQ AGN, i.e., star formation (SF), accretion disc winds,  coronal disc emission, low-power jets\footnote{Here we use the 'jet' term to indicate an outflow that becomes collimated, whereas a 'wind' is an uncollimated outflow. Both transport outwards mass, energy and angular momentum.} or a combination of them (Fig.~\ref{fig:sketch}). Indeed, the relatively few available studies of RQ AGN at high and low resolution (e.g., \citealt{kellermann89,kellermann94,barvainis96,kukula98,ulvestad05,leipski06,padovani11,doi11,Zakamska_2016,zakamska16b,hartley19}) have generally led to mixed results. This is not surprising, as all RQ AGN are powered by a SMBH and emit X-rays (from a corona), most are in star-forming hosts, and many display outflows.

In this review we discuss, using an observational approach, the physical mechanisms that in our opinion may significantly contribute to the radio emission in RQ AGN (Section 2), highlighting analogies and differences between other radio-emitting astrophysical accreting systems (Section 3). In Section 4 we provide testable theoretical predictions to evaluate the dominant mechanism and in Section 5 we draw the required recipes for current and future radio facilities; we conclude with Section 6. We refer in general to RQ AGN with radio absolute weak powers and/or radio faint emission with respect to the emission at higher frequencies, including in this class RQ QSOs (RQQs), Seyfert galaxies and LINERs\footnote{In this work we do not enter into the details of the different RQ AGN classes but instead provide a broad view of radio emission in (mostly) local RQ AGN. For a recent and detailed discussion of AGN classification and an alternative naming of RL/RQ AGN as ``jetted'' and ``non-jetted'' AGN see \cite{Padovani_etal_2017} and \cite{Padovani_2016,Padovani_2017}.}. Our purpose in this review is to alert the general AGN community that the radio regime is becoming a powerful new tool for studies of
a variety of physical mechanisms in RQ AGN, at unprecedented resolution.

\section{Physical and radiative mechanisms}

\begin{figure*}
\centering
  \includegraphics[width=0.8\linewidth]{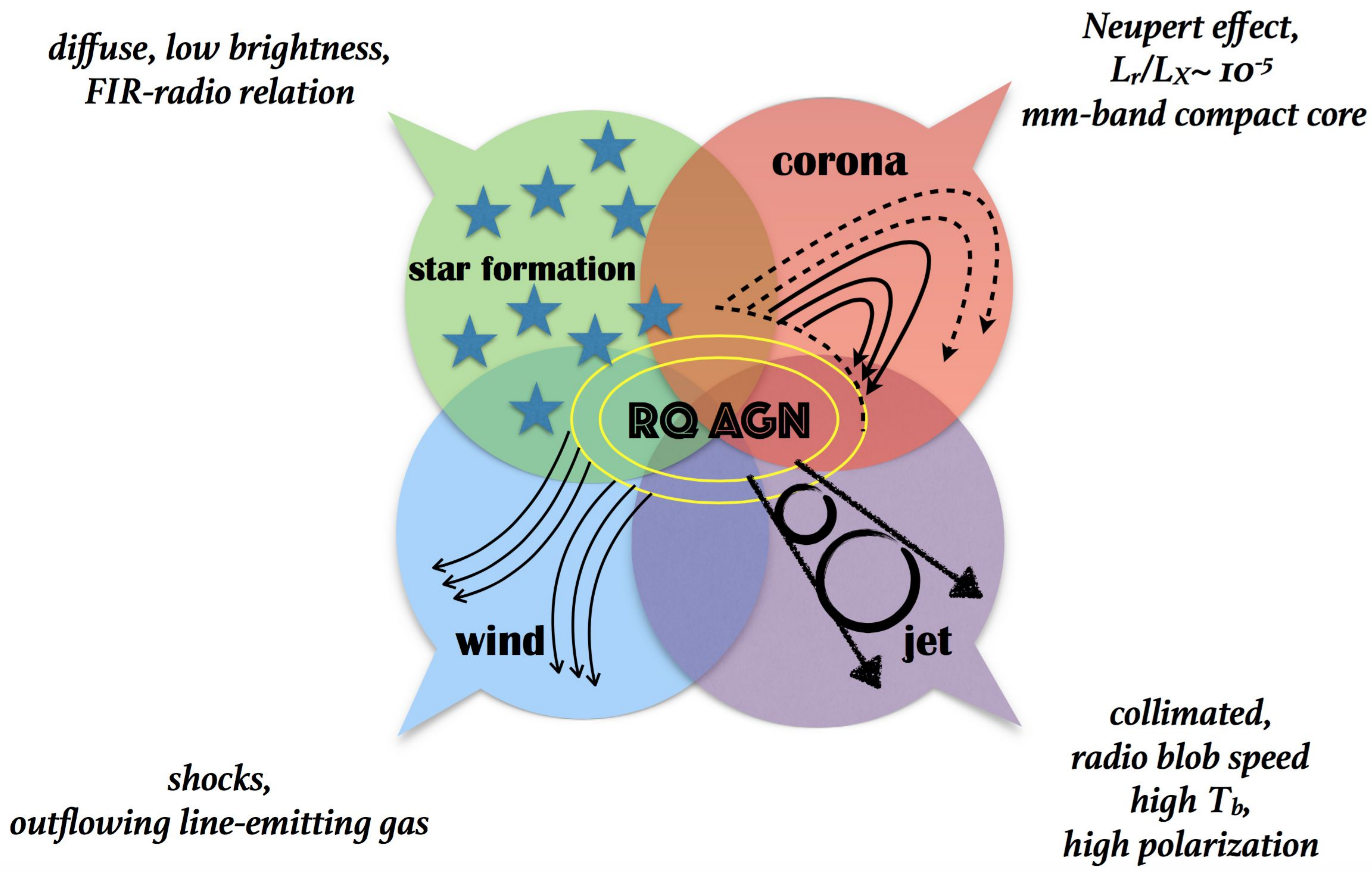}
  \caption{Sketch of the four main different physical mechanisms discussed in this work, to account for the origin of the radio emission in RQ AGN. For SF a diffuse low-surface brightness extended emission is expected, by tracing the host morphology, in significant excess with respect the FIR-radio correlation valid for quiescent star-forming galaxies \citep{Sargent_2010}; the coronal emission would emerge from the optically-thin radio emission in case of a high radio frequency flat-spectrum unresolved core with a radio and X-ray luminosities consistent with the \citet{guedel93} relation (L$_{R}$/L$_{X}$ $\sim$ 10$^{-5}$) and a specific dependence of the correlated radio--X-ray light curves, as predicted by the Neupert effect \citep{neupert68}; the measurements of the jet blob speeds, a high brightness temperature (T$_{b}$ $>$ 10$^{8}$ K) and high polarization are key parameters to identify a collimated jet emission; for a wind origin, shocks on the interstellar medium are expected to ionize the gas and emit blue-shifted X-ray, optical, and molecular lines, sign of the outflowing material.}
  \label{fig:sketch}
\end{figure*}

Multiple physical radio-emitting processes are thought to be in act in the nuclear regions of RQ AGN (see  Fig.~\ref{fig:sketch} for a simple sketch). Below we discuss the observational evidence and interpretation related to each mechanism separately.

\subsection{Jets}\label{sec:jets}

Ejected material from an accreting SMBH can take the form of highly collimated and powerful relativistic jets \citep{begelman84}, from pc to Mpc scales, emitting synchrotron emission and up-scattered photons via inverse Compton into X-rays and $\gamma$-rays. AGN jets display a broad variety of radio structures (for example FR~I and FR~II, \citealt{fanaroff74}), with a wide span of intrinsic powers. 

Flat or slightly inverted radio spectra from very compact sources are generally associated with an optically thick launching region, such as a jet base \citep{blandford79,reynolds82}\footnote{Note that jet bases do not always have flat spectra, as in the case of some FR~I sources \citep{laing2013}.}, obtained from the overlap of magnetised self-absorbed synchrotron blobs or multiple components with an optically-thin low energy turnover \citep{falcke95}. The high-brightness temperature of radio cores is thought to be the result of non-thermal processes from relativistic electrons \citep{blundell98}. The size of the visible inner AGN jet cone scales proportionally with wavelength and is typically on the pc scale. 

At low radio luminosities, follow-up radio surveys of optically selected nearby low luminosity AGN (LLAGN, with luminosities at 1.4~GHz below 10$^{42}$ erg s$^{-1}$, including LINERS and low-luminosity Seyferts (\citealt{falcke00,ho01,nagar02b,nyland16,baldi18,chiaraluce19}) have yielded extremely high detection rates, with the radio emission having predominantly a flat spectrum, compact core morphology, occasionally accompanied by jet-like features on pc scales with intermediate resolution ($1-0.1 \arcsec$ with VLA and eMERLIN). The radio emission on mas scales is often compact and most of the large scale radio emission is resolved out when imaged with the long-baselined arrays \citep{falcke00, nagar01}. 
In particular, Seyferts often display compact nuclear radio emission, occasionally similar to the jetted LINERs (see left panel of Fig~\ref{fig:examples}) and more frequently similar to low-brightness diffuse lobed structures, indicative of sub-relativistic bulk speeds  (e.g., \citealt{baum93,kukula96,gallimore06,kharb14,singh15}), as measured on pc scales in a few Seyferts \citep{roy00,middelberg04}. 
Long-baseline radio observations show that most Seyfert galaxies possess compact sub-pc scale nuclear emission with brightness temperature $>$10$^{7}$ K and occasionally misaligned pc-scale jets, possibly suggestive of jet precession \citep{middelberg04,giroletti09,panessa13,kharb18}. 

Emission from RQ AGN has been sometimes explained with a scaled-down version of more powerful jets, where the difference between RL and RQ could be due to a different efficiency in accelerating relativistic electrons on the sub-pc scale and in collimating the flow \citep{falcke95}. The jet properties likely evolve with the bolometric AGN luminosity. This results in a higher fraction of radio jets observed in local low-luminosity RQ AGN than in powerful quasars \citep{blundell01,ho01,ulvestad01,blundell03,heywood07,baldi18}, although selection effects may play a role. 
Radio structures of RQQ are found to be larger than those of Seyfert galaxies, but with similar morphologies \citep{leipski06}. Similarly, LINERs typically show core-brightened radio jets reminiscent of those observed in small FR~Is \citep{capetti17}.

\subsection{AGN winds and thermal free-free emission}\label{sec:outflows} 

\begin{figure}
\begin{center}
  \includegraphics[width=1\linewidth]{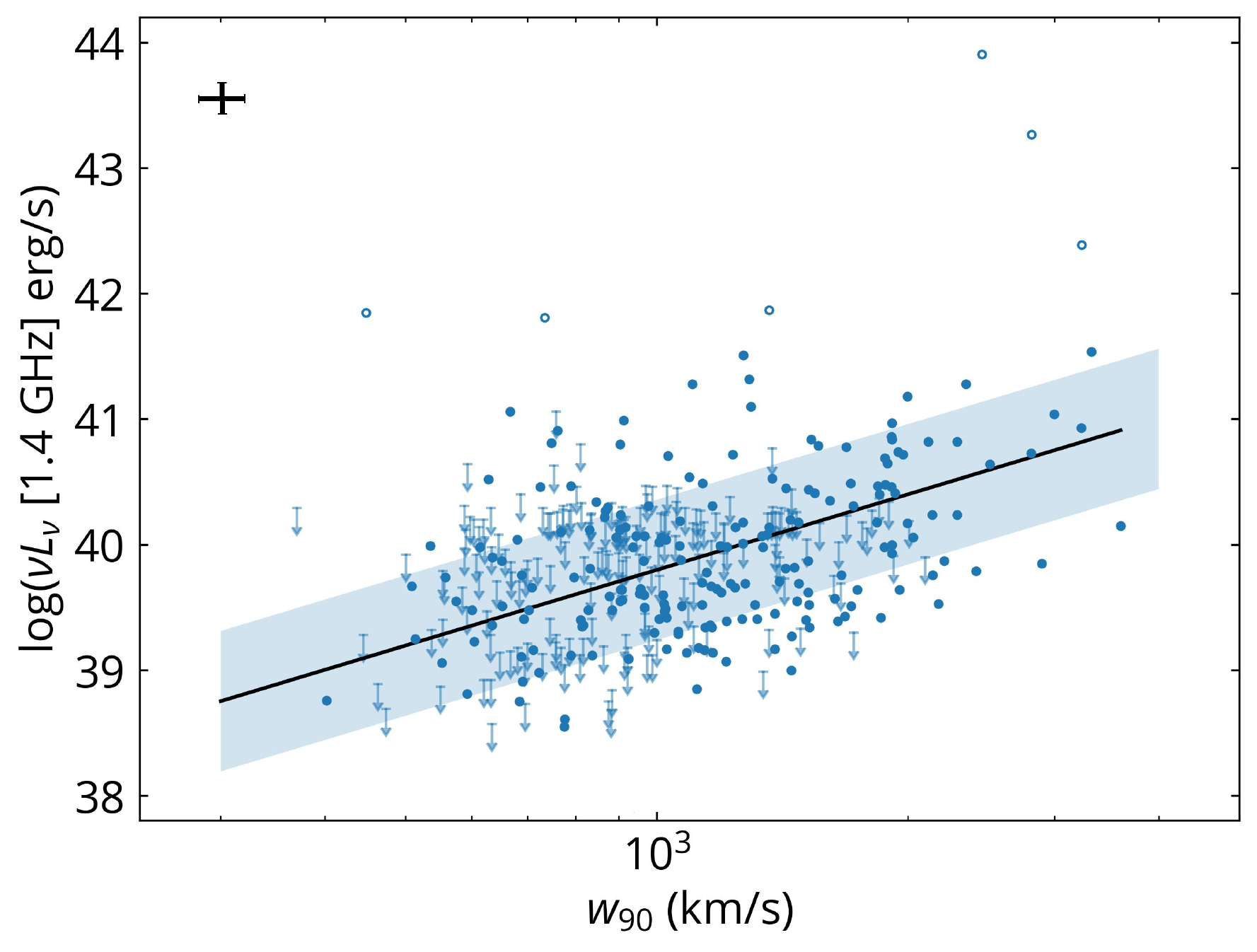}
  \caption{Radio luminosities at 1.4 GHz versus the velocity width containing 90 per cent of [O~III] line for the low-redshift (z$<$0.8) type-2 quasars \citep{zakamska14}, with radio luminosity upper limits from FIRST and NVSS surveys. The error bar at the top-left corner describes the typical uncertainties of the data. The black line is the quadratic fit on the low-z RQ sample and the shaded region represents the root-mean-square scatter. Figure adapted from Fig. 7 in \citet{hwang18} (courtesy of Nadia Zakamska).}
\label{fig:Zakamska}
\end{center}
\end{figure}

A rich and complex phenomenology of galactic winds located at different spatial scales has been revealed by multi-wavelength observations. Sub-relativistic ($0.1-0.25$\,c) ultra-fast outflows (UFOs) in the vicinity of the BH (gravitational radius $R_g$ scales) have been typically observed as Fe K-shell X-ray absorption-line features \citep{chartas09,tombesi10,pounds14,nardini15}, as well as fast multiple components in soft X-ray grating spectra \citep{gupta15,longinotti15}. Lower velocities and temperatures from photo-ionised winds are detected in soft X-rays and UV spectra \citep{crenshaw99, laor02}. On kpc scales, broad and high velocity kinematic components are present in Seyferts and more luminous AGN as seen in [O~III]$\lambda$5007 emission line profiles \citep{mullaney13,zakamska14} as well as in molecular outflows \citep{morganti15}, occasionally associated with diffuse radio emission (see middle panel of Fig~\ref{fig:examples}). 
It has been suggested that compact radio emission plays an important role in disturbing the [O~III] emitting gas in optically selected AGN \citep{capetti99,mullaney13,harrison15}. The correlation found in RQ AGN between radio emission and [O~III] line velocity has been explained as the effect of wind shocks 
(Fig.~\ref{fig:Zakamska}, \citealt{zakamska14}), with an extension to the most luminous high-redshift quasars, as discussed by \citet{hwang18}. 
Such shocks accelerate relativistic electrons producing synchrotron radio emission on scales $> 100$ pc, with powers at the level of those observed in RQ AGN, $\nu$L$_{\nu}$ $\sim$ 10$^{-5}$ L$_{\rm AGN}$ \citep{Nims_2015}. In low-luminosity Seyferts, the base of the outflow is found to be coincident with the unresolved nucleus, pointing to the AGN as the predominant ionising source of the outflowing gas \citep{lena15}. Indeed, low-brightness temperature ($T < 10^{5}$ K) VLBI radio cores are consistent with an extended thermal origin from a small scale outflow \citep{christopoulou97}. In addition, the positional coincidence with low-luminosity water masers suggests that the radio continuum may arise from the inner regions of a molecular disc or from a nuclear wind \citep{gallimore04,hagiwara07}.

Disc winds may also produce radio emission via brems\-strahlung free-free processes from an optically thin ionized plasma \citep{blundell07, blustin09, steenbrugge10}. It has been shown that for a sample of quasars, free-free emission from a disc wind cannot mutually account for the observed radio and X-ray luminosities \citep{steenbrugge11}. However, simple emission measure estimates from the narrow line region gas indicate that the free-free emission of this gas may lead to detectable fluxes in the millimetre (mm) band (Baskin \& Laor, in preparation). 

Competing driving mechanisms have been proposed for AGN winds, such as starburst outflows \citep{heckman15} and magnetic fields \citep{fukumura10}.

Supernovae, which are common in intense SF regions, can blow super-winds, which inflate radio bubbles. The energetic budget of these super-winds \citep{heckman15} from high star formation rates ($\sim$30-50 M$_{\odot}$ yr$^{-1}$) appears to be able to account for the global properties of the ISM and the extended radio structures observed in several Seyferts and starburst galaxies \citep{veilleux94,genzel95,maiolino98,kharb06}. 

Magneto-hydrodynamic (MHD) accretion-disc winds have been proposed to explain the wide variety of X-ray winds \citep{fukumura14}. 
Interestingly, UFOs have been observed both in RQ and RL AGN (e.g., \citealt{tombesi11,longinotti15}), suggesting that jets and winds may co-exist \citep{giroletti17}, contrary to BH X-ray Binaries (XRBs), where the presence of an accretion disc wind is observed only during the softer X-ray states (see Section~\ref{comparison}) and could be responsible for the quenching of the radio jet (\citealt{ponti12}, see also NS binaries \citealt{diaz16}). On the theoretical side, MHD models predict the possible co-existence of jets and winds, suggesting that an axial collimated jet of low matter density and an extended disk wind maybe both part of the same outflow (e.g., \citealt{ferreira10}). The former is likely connected to the BH, while the latter arises from the inner regions of the accretion flow \citep{sasha11}.

\subsection{Accretion Disc Coronae}
\label{sec:corona}

The presence of a hot $T \sim10^{9}$\,K corona in the vicinity of BH accretion discs has been suggested as early as fifty years ago to explain the non-thermal (Comptonised) X-ray spectrum, first for stellar BHs \citep{shapiro76} and later for AGN \citep{haardt91}. Compared to other potential radio sources in the AGN and host galaxy, the coronal component comes from the smallest region around the BH, i.e., 10--10,000 $R_g$ (about a milli-pc for a BH of $10^8$ M$_{\odot}$). Consequently, it is largely unresolved by contemporary instruments, and its size needs to be assessed from models or variability, as discussed below.

A first hint that radio emission from RQ AGN is related to the central nuclear source comes from the correlation of the radio luminosity $L_R$ and the X-ray luminosity $L_X$ \citep{brinkmann00,salvato04,wang06,panessa07,panessa13,panessa15}. The strongest evidence comes from the Palomar-Green (PG) quasar sample ($z < 0.5$, \citealt{bg92}), where it was found \citep{laor08} that $L_R$ at 5\,GHz and $L_X$ (0.2 -- 20 keV) are not only correlated over a large range of AGN luminosity, but follow the well established relation for coronally active cool stars of $L_R / L_X \sim 10^{-5}$ \citep{guedel93}. The origin of this relation is not well understood from first principles in stars either, but it may suggest a coronal, magnetic origin for the radio emission in RQ AGN from magnetic reconnection phenomena.

The size in pc, $R_\mathrm{pc}$, of a self-absorbed synchrotron source, assuming a circular optically thick spot on the sky, decreases with frequency according to the equation derived from \citet{laor08} (eq. 19 therein):  
\begin{equation}
R_\mathrm{pc} \simeq 0.54L^{1/2}_{39}\nu^{-7/4}_\mathrm{GHz}B^{1/4} 
\label{Rpc}
\end{equation}

\begin{figure}
\begin{center}
  \includegraphics[width=0.95\linewidth,height=0.45\linewidth]{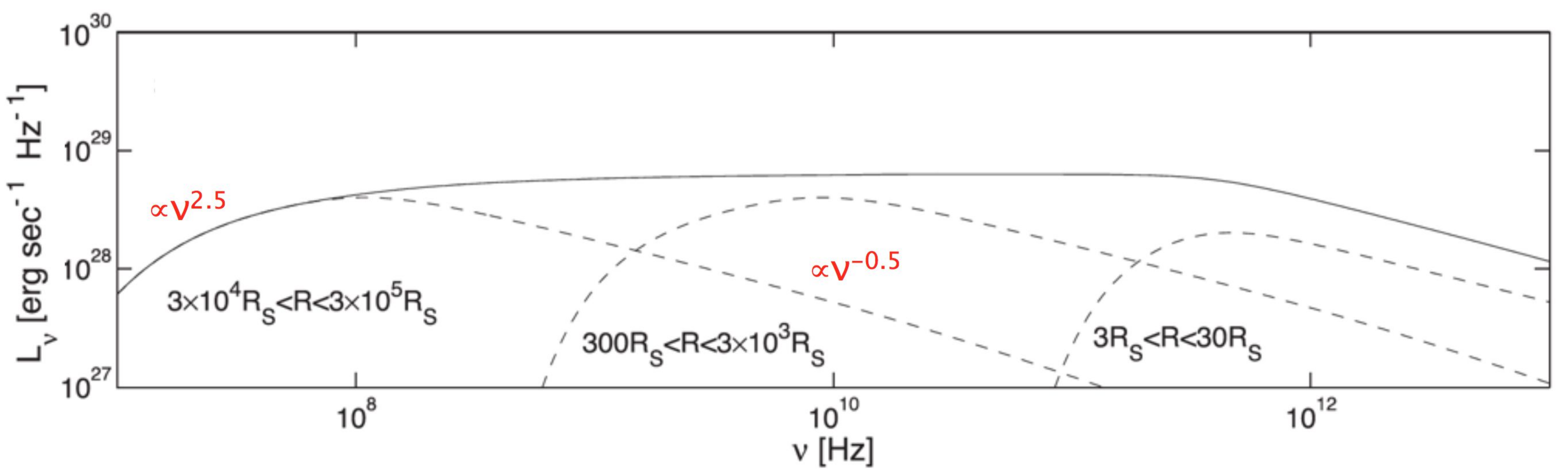}
  \caption{The spectrum of a distribution of power-law electrons
in a spherical corona. The solid line represents the overall
emission, and the dashed lines the contributions of the different spherical components at different gravitational radii. The specific intensity increases with the
frequency as $\propto \nu^{2.5}$ in the optically thick regime at low frequency and decrease as $\propto \nu^{-0.5}$ in the optically thin regime at high frequency. Figure adapted from Fig. 4 of \citet{raginski16}.}
  \label{fig:corona}
\end{center}
\end{figure}

\noindent Hence, the 5\,GHz emission likely comes from the broad line region scale, of about $10^4 R_g$. Observing the radio corona at scales comparable to those of the X-ray source ($\sim 10~R_g$) requires observing at much higher frequencies (\citealt{behar15}) in the mm band at $100 - 300$\,GHz. At higher frequencies, cold dust emission from the host dominates. At 100 GHz (3 mm), flux densities have been measured for a few dozen local AGNs \citep{behar15,behar18}, and found to exceed the interpolation of the low-frequency steep slope power-law. Source sizes were estimated from Eq. (1) to be of less than a light day, and between 10 and 1000 Rg, which is comparable to the variability size of the X-ray source. Detections of mm excess in Seyfert galaxies confirm the interpretation of a compact, optically thick core that is superimposed on the steep power law of more extended structures that dominate at lower frequencies \citep{inoue14,doi16,inoue18}. Recently, \citet{inoue18} reported the detection of coronal synchrotron radio emission from two nearby Seyfert galaxies associated with a magnetic field of $\sim10$ Gauss on scales of $\sim$ 40 $R_g$ from the central BHs. 
Theoretical computations \citep{raginski16} explored this conjecture quantitatively showing that an X-ray corona can produce a flat synchrotron radio spectrum up to a break at 300~GHz or higher (Fig.~\ref{fig:corona}).

The smoking gun of the coronal origin of the radio emission in RQ AGN would be the detection of the radio and X-ray variability correlation observed in stellar coronae, the so-called Neupert effect \citep{neupert68,gudel02} in which the radio flare precedes the X-ray one, and over the duration of the flare the time derivative of the X-ray light curve varies in a fashion similar to the radio light curve $L_R \propto dL_X/dt$. This is interpreted as a result of energetic electrons that emit synchrotron radio ($\propto L_R$) depositing their heat in the corona that subsequently shines out the {\it total} energy in X-rays, thus $L_X \propto \int{L_Rdt}$, and hence the observed correlation. 

Flares and dips in the hard (jetted) state of X-ray binaries have shown a related effect of $L_{opt} \sim -dL_X/dt$, where $L_{opt}$ is the optical luminosity coming from synchrotron emission \citep{malzac03}, but where the minus sign makes it different from the Neupert effect. Indeed, \citet{malzac04} interpret the effect in terms of a jet-disc coupling and a common energy source for the synchrotron jet (optical emission) and the Comptonizing electrons (X-rays), so that one is enhanced at the expense of the other and vice versa. The analogy to AGNs is unclear yet intriguing, and should be studied further as more multi-wavelength observations of RQ AGN become available (see Section 3).

\subsection{Star formation}\label{sec:SF}

SF is the process by which dense regions within 
molecular clouds in the ISM collapse and form stars. SF activity produce thermal and non-thermal radio emission thanks to  the presence of strong magnetic fields and hot plasma from supernova remnants and cosmic rays. Star-forming regions are typically host-like extended, diffuse, clumpy and with low surface brightness (see a clear example in the right panel of Fig.~\ref{fig:examples} and others in \citealt{Orienti_2015}). 
The SF phase, in particular, is  characterised by steep GHz spectra ($\alpha_{\rm r} \approx 0.7$) dominated by 
synchrotron emission at low frequencies with a flat free-free component, which  becomes predominant at $\nu \gtrsim 30$ GHz \citep{Condon_1992}. 

Far-Infrared (FIR) emission, indicator of dust and cold gas, an ideal cradle for new generations of young stellar populations, has been found to correlate with radio emission for Seyferts and relatively low-redshift RQQ, which follow the so-called FIR -- radio correlation typical of star-forming galaxies (SFGs) \citep{Sopp_1991, Sargent_2010}, which is understood to be driven by recent SF. In addition, the FIR flux density in Seyfert galaxies correlates better with the low-resolution kpc-scale radio  flux density rather than with the high-resolution pc-scale emission \citep{Thean_2001}, which points to an SF origin for the former at the galaxy scale. All of the above fits with the fact that low-luminosity RQ AGN are usually hosted in late-type galaxies, which are commonly star-forming.

\begin{figure}
\begin{center}
  \includegraphics[width=1\linewidth]{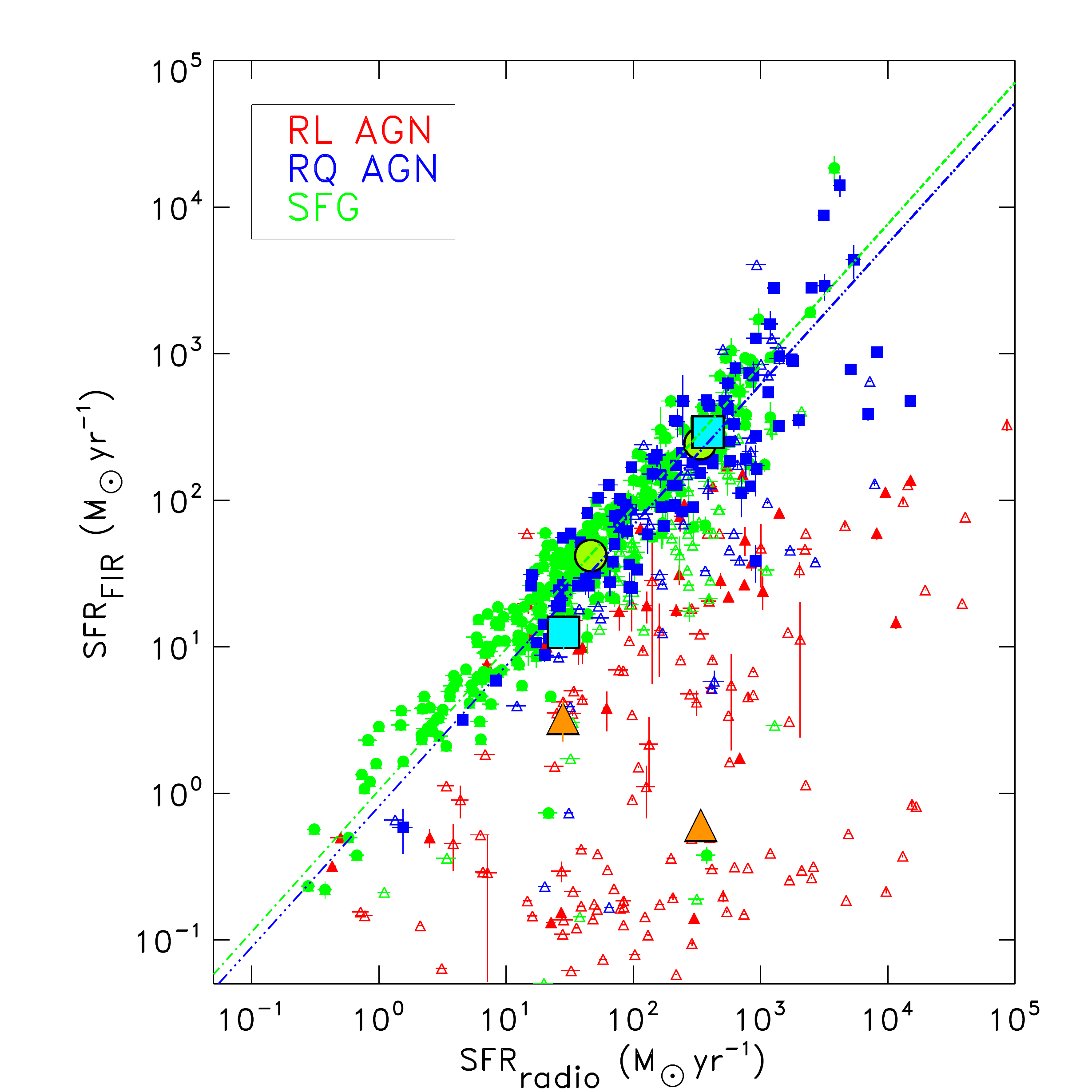}
  \caption{SF rate  derived from the FIR luminosity versus the SF rate from $P_{\rm
    1.4GHz}$ for the E-CDFS sample. SFGs are plotted as green circles, RQ
  AGN as blue squares, and RL AGN as red triangles. Full symbols represent
  sources detected in at least one Photoconductor Array Camera and Spectrometer 
  (PACS) filter, while sources shown as
  empty symbols are \textit{Herschel} non-detections, for which SFR$_{\rm
    FIR}$ is less robust.  Large symbols with lighter colours are the
  results of a stacking analysis. The two lines are the best fits for the
  SFGs and RQ AGN with PACS detection. Error bars denote the uncertainties on the derived SF rates. Figure reproduced from
  \cite{Bonzini_2015}, Fig. 3, with permission.}
  \label{fig:FIR-Radio}
\end{center}
\end{figure}

The situation at higher powers (and redshifts), however, is not so clear-cut, as discussed by \cite{Padovani_2016}. On the one hand, it appears that the main contribution to  radio emission in sub-mJy RQ AGN selected at 1.4 GHz at $z \sim1.5$ comes from  SF activity in the host rather than from the AGN \citep[e.g.,][] {Bonzini_2015,Padovani_2016,DelVecchio_2017,gurkan18}. This scenario is based on the fact that the SF rate derived from the FIR luminosity as traced by {\it Herschel}  agrees extremely well with that estimated from the radio power under the assumption that the latter is due to SF \citep[][see Fig.~\ref{fig:FIR-Radio}]{Bonzini_2015}. This is further supported by: i) the fact that Extended {\it Chandra Deep Field}-South (E-CDFS) RQ AGN occupy the same locus as SFGs also 
in the SF rate (SFR) -- stellar mass plane \citep[Fig. 5 of][]{Bonzini_2015}; ii) the similar radio luminosity evolution of E-CDFS RQ AGN and SFGs \citep{Padovani_2016}; iii) the $L_R / L_X$ ratios of E-CDFS RQ AGN, which show higher values than those expected for coronal emission (Sect. \ref{sec:corona}) and synchrotron emission from AGN outflows \citep[][and Sect. \ref{sec:outflows}]{Nims_2015}, but smaller than the corresponding ratios for relativistic jets. Moreover, the radio sizes of radio-selected AGN with $\rm SFR_{\rm radio} \approx\rm SFR_{\rm optical-FIR}$ are typically a factor  of 2 larger than those for which $\rm SFR_{\rm radio}$ exceeds by more than $3\sigma$ $\rm SFR_{\rm optical-FIR}$ \citep{Bondi_2018}. This is fully consistent with SF-related radio emission in the former sources. A core, possibly BH-related component, however, can also be present in radio-selected RQ AGN, which clearly show extended SF in other bands (see also Sect. \ref{sec:jets} for local sources). Some sub-mJy RQ AGN, in fact, show evidence 
for relatively strong compact radio cores \citep[e.g.,][]{Chi_2013,Maini_2016,Herrera_2016}, which 
suggests that the AGN component might be at the
same level as, or even stronger than, the SF one. In addition, SFGs can hide LLAGN at their centre able to launch pc-scale radio structures \citep{baldi18}. At higher resolutions, one needs to keep in mind, though, that VLBI detections are still biased towards AGN-dominated sources, resulting in an incomplete view of the SF at the nuclear scale.

On the other hand, different results can be obtained for RQ AGN samples selected in other bands (see also Sect. \ref{sec:corona} for the PG sample). 
\cite{Zakamska_2016} have come to the conclusion that radio emission in their 
sample (optically selected, $S_{\rm 1.4 GHz} > 1$ mJy, $z < 0.8$, and $L_{\rm bol} > 10^{45}$ erg s$^{-1}$) is dominated by quasar activity, not by the host galaxy
(see also \citealt{White_2017}). However, \cite{Kellermann_2016}, by studying 
$0.2 < z < 0.3$ optically selected quasars at 6 GHz, agree with the results obtained from the E-CDFS sample. One might think that  
power might play a role, since E-CDFS RQ AGN have relatively
low luminosities ($\langle L_{\rm x} \rangle \sim 10^{43}$ erg s$^{-1}$
i.e., $L_{\rm bol} \approx 3 \times 10^{44}$ erg s$^{-1}$; see also  
\citealt{Rosario_2013}, who find similar results to \citealt{Bonzini_2015}).
Other complicating issues might include different definitions of RQ/RL AGN, 
and the radio frequency at which the studies are made\footnote{At low frequencies  
the steep synchrotron SF-related emission dominates, while at 
higher frequencies a flat component from a radio core or a corona should be relevant  (Sects. \ref{sec:jets} and \ref{sec:corona}; but note that the \citealt{Zakamska_2016} results were obtained at 1.4 GHz, as the
E-CDFS ones; and, moreover, as discussed above, at high frequencies a flat component is expected also from SF).}. 

\section{Comparison with other accreting sources}
\label{comparison}

A possible pathway towards the understanding of the mechanisms producing radio emission in RQ AGN is through the comparison with other accreting sources, from Young Stellar Objects (YSOs), to XRBs (either BHs or neutron stars) and active SMBHs. These various accreting sources display substantial differences, e.g., in the central mass, outflow velocities, disc temperatures, energy output and magnetic fields. However, a number of common features and parameters (e.g., disc morphology, degree of collimation, ejection velocity and direction) has led to the suggestion that they may all share similar physics in the phenomenon of accretion and ejection \citep{price03}.

\begin{figure}
  \includegraphics[width=1\linewidth]{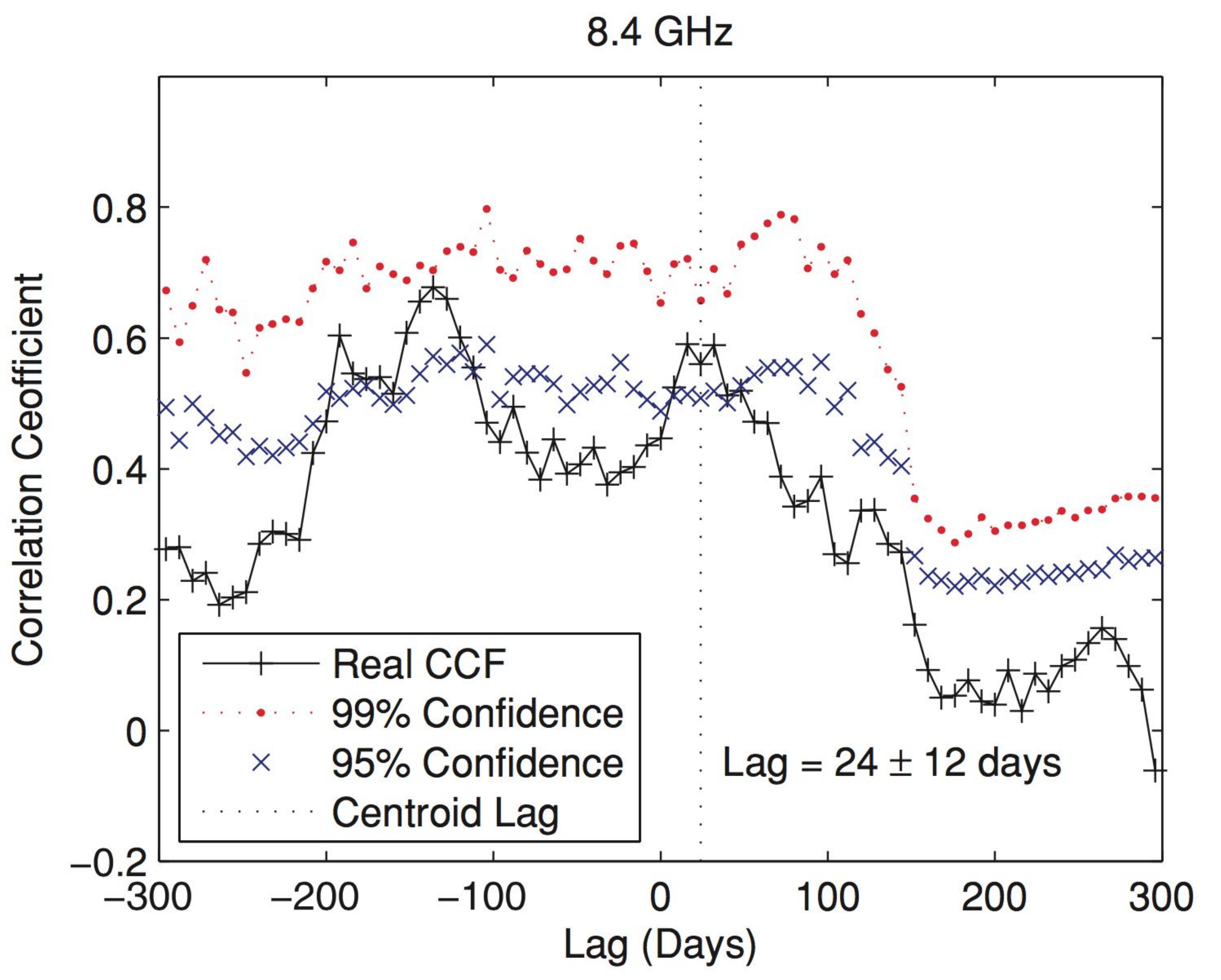}
  \caption{Discrete cross-correlation function  plot showing  time-lag  against  the  correlation  coefficient  for X-ray to radio (8.4 GHz lagging X-ray) (black crosses) for NGC 7213. A time lag of 24$\pm$12 d is found to maximise the cross-correlation function. The 99 and 95 per cent local significance confidence levels are plotted (reproduced Fig. 2 from \citealt{bell11}).}
  \label{fig:bell11}
\end{figure}

From XRBs we have learnt that jets and outflows are intermittent in their nature, likely following accretion state changes \citep{fender04,ponti12}. Different attempts have been made to unify XRBs with AGN \citep{falcke04,kording06}. More specifically, XRBs in their low-accretion rate hard state, can launch jets (with moderately relativistic speeds), similar to AGN. In this state, the observed relationship between their X-ray and radio luminosities \citep{corbel03,fender04} provide evidence for a physical connection between the inflow and the collimated outflow. Conversely, in the higher accretion rate 'soft state' no evident radio activity has yet been found despite very deep searches \citep{russell11}. Strong radio flares are often detected through the transition between the two states at high accretion rates, in some cases resolved into discrete symmetric blobs of material. A complex radio X-ray relationship can be observed during these transitions, involving an X-ray dip preceding a radio flare \citep{belloni97,mirabel98}, similarly to what is seen in the radio galaxy 3C~120 \citep{marscher02,chatterjee09}.

RQ AGN have shown cm/mm-band variability on day-month-year timescales (e.g., \citealt{wrobel00, barvainis05,chatterjee09,mundell09, baldi15}), which, at face value, appears to be similar to the behaviour of binaries in their hard states. Nevertheless, simultaneous radio/X-ray monitorings are rare and have so far yielded only weak or no correlations, with ambiguous results (see Figure \ref{fig:bell11}, \citealt{bell11,king11,king13,jones11,jones17}). 

In the light of a possible unification of radio AGN with low/hard state XRBs, relevant models relating jet radio luminosity to the BH mass and accretion rate were proposed \citep{markoff01,yuan02,heinz03} to account for the accretion-ejection phenomenon in compact objects. Further observational support for such models is the so-called 'fundamental plane' (FP) of BH activity, relating mass, radio luminosity (at 5 GHz, 6 cm) and X-ray luminosity (e.g., \citealt{merloni03,falcke04}) which unifies XRBs, RL and RQ AGN. The scatter around the plane is substantial but is reduced if only low accretion rate BHs are considered (e.g., \citealt{kording06,falcke04,plotkin12}). 

\begin{figure}
\hspace*{-0.7cm}
  \includegraphics[width=0.8\linewidth,angle=90]{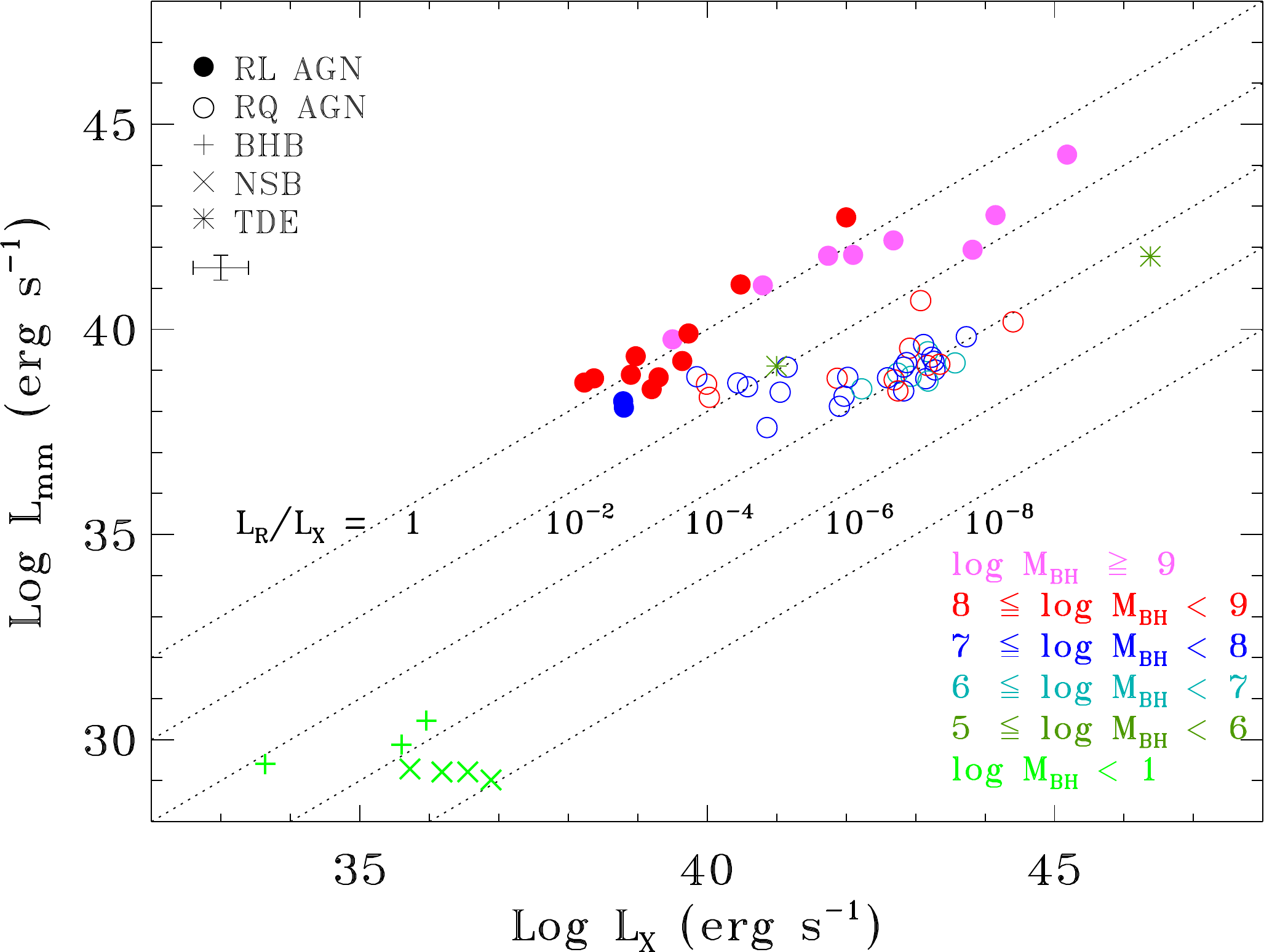}
  \vspace{0cm}
  \caption{X-ray luminosity (2-10 keV, erg $s^{-1}$) versus mm-band luminosity (90-100 GHz, erg $s^{-1}$) for RL AGN \citep{hardcastle08,cotton09,doi11}, RQ AGN \citep{doi11,behar15,behar18}, XRBs (BH and neutron star binaries, \citealt{fender00,berger12,yuan16,diaz17,diaz18,tetarenko18a,tetarenko18b}) and TDEs (IGR~J12580+0134 and SWIFT~J1644+5734). We colour  the symbols based on the BH masses (M$_{\odot}$) derived from stellar velocity dispersions. For AGN the observations in the two bands are not simultaneous, while for the other classes of sources, which are known to be transients, the data are taken on the same day or a few days apart. Precisely, we include XRBs (V404 Cyg, Cyg X-1, Aql X-1, 4U~1820-30, and MAXI~J1820+070) in hard states and transitioning to soft states. The dashed lines correspond to L$_{mm}$/L$_{X}$ of 1, 10$^{-2}$, 10$^{-4}$, 10$^{-6}$, and 10$^{-8}$. In the upper-left corner, the plot legend includes the typical error bar associated with the data-points. Figure updated and adapted from Fig. 3 of \citet{behar15}.}
  \label{mm-x}
\end{figure}

The wealth of incredible details known on jets and outflows structures and chemistry in YSOs could ease our comprehension of such phenomena in all accreting systems. Indeed, most jet parameters are derived from imaging and spectroscopy (e.g., velocity, mass loss rate, composition) and information on 
strength and direction of the magnetic fields would provide an almost complete characterization of outflows in YSOs. However, the radio emission observed is mainly non-thermal \citep{tobin15}. In this respect, YSOs and AGN complementary studies are of high value, as magnetic fields in AGN can be tested via polarisation. Similarities can be drawn between YSOs and both AGN and XRBs, such as the detection of proper motions or 
the birth of new radio knots ejected in this case by the star (e.g., \citealt{osorio17}). In addition, in YSOs the accretion is episodic with periods of quiescence as suggested by the outflow structures \citep{bally16} and strictly connected to the ejection flow.

Hints of disc/jet coupling are found also in weakly accreting sources such as cataclysmic variables \citep{kording08}, as well as in highly accreting systems such as Ultra Luminous X-ray sources \citep{roberts07,cseh15,vandeneijende18} and Tidal Disruption Events (TDEs, \citealt{dai18}). In particular, TDEs offer the opportunity to observe in real time the on-set of jet activity at high \citep{bloom11} and low power \citep{pasham18, vanvelzen16, alexander16, alexander17, mattila18}. The detection of a cross-correlation between the X-ray and radio light curves in the case of the TDE ASASSN-14li suggests that the soft X-ray emitting disc regulates the radio emission, providing evidence for disc-jet coupling also in recently formed jets \citep{pasham18}.

As an attempt to unify the different accreting classes, in Figure~\ref{mm-x} we plot the mm-band luminosity ($\sim$3 mm) versus the X-ray luminosity (2--10 keV) for RQ and RL AGN, XRBs (BHs and neutron stars), and TDEs, coloured in bins of BH masses. Since the mm-band emission is expected to come from inner regions with respect to its cm-band counterpart (by a factor $\sim$200, see Eq. 1), the mm/X-ray luminosity ratio should be more representative of the properties of accretion and ejection in compact objects. This plot is only exploratory as mm-band observations are available only for a few sources so far and objects in different accretion states are considered.
However, there is a clear displacement between RL and RQ AGN by 3-4 orders of magnitude at a given L$_{\rm X}$. Furthermore, a dependence of the mm/X-ray ratio on BH mass is also evident, with the small BH masses showing lower ratios, as also found in the FP \citep{merloni03}. While TDEs shows mm/X-ray ratios similar to AGN, XRBs have even lower ratios, probably due to their smaller masses (note that several uncertainties are associated with the extremely object-dependent behaviour of XRBs).

\section{Testable theoretical predictions}\label{sec:predictions}
 
\begin{figure*}
  \includegraphics[width=1\textwidth]{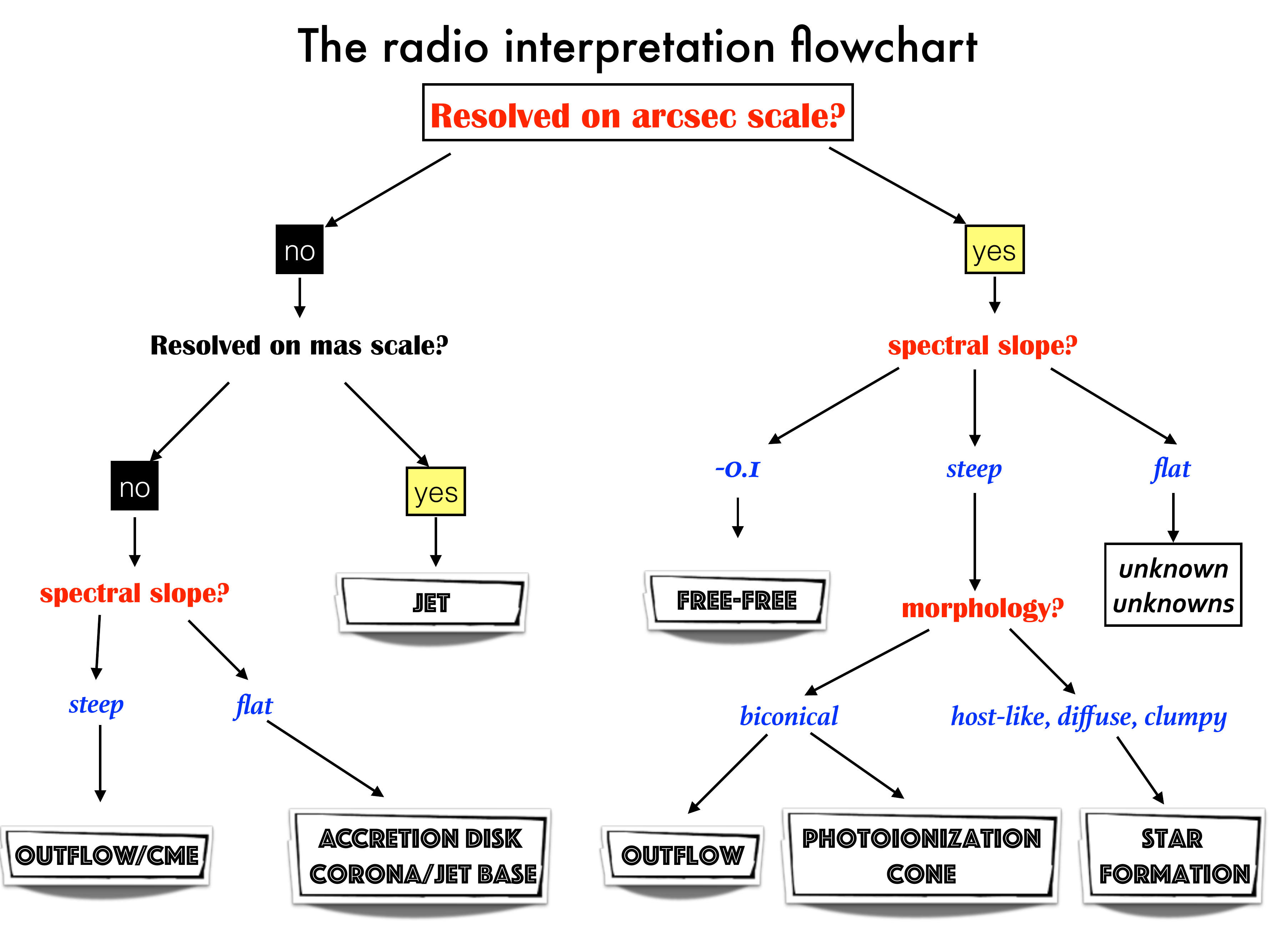}
  \caption{Flow chart to guide the interpretation of the radio emission in local RQ AGN. The size of the resolved radio emission at different resolution (arcsec and mas scales corresponds to kpc and pc scales, respectively for local galaxies), the surface brightness of the emission and the typical spectral slope derived between 1.4 and 8.5 GHz can be used as key parameters to approximately identify the different physical process involved in the radio band.}
  \label{fig:flow}
\end{figure*}

In Figure \ref{fig:flow}, we present a scheme that could serve as a guide to evaluate which is the dominant physical mechanism in a RQ AGN radio map, by means of its morphology and spectral slope. Testable predictions can be drawn from simple arguments linked to the examined physical processes in this work and based on their main characteristics (see Figure~\ref{fig:sketch}).

In case of a jet origin, the physical mechanism may be similar to that of a RL jet ejection process, but scaled-down in luminosity by a factor of $\sim 10^3$. High resolution (i.e. with VLBI) observations may resolve the mas-scale jet structure and measure high-brightness knots motions for a larger number of sources viewed face-on, while for those viewed pole on, the emission may remain unresolved also on mas scale. Fundamental properties of the non-thermal emission component can be derived via radio observations, such as an estimate of the magnetic field (e.g., from frequency break in the spectra such as in \citealt{bontempi12} or spectral energy distribution fitting as in \citealt{inoue18}) and the degree of polarisation \citep{giroletti05}.

Extended diffuse radio emission may also be due to a wind-induced shock. Since AGN emission is non-isotropic, the wind is also expected to show some conical or double-lobed structure, in contrast with the host SF. Such a wind will inevitably manifest itself through the kinematics of spatially extended line emission \citep{veilleux05} with a potentially strong impact on the galaxy \citep{harrison18}. The advantage of the radio band is that it allows us to probe the wind well down to sub-arcsec scale, otherwise accessible only with optical-IR space telescopes and interferometers or integral-field spectrographs.

Thermal free-free emission is simple to identify through its flat spectral slope of $L_{\nu}\propto \nu^{-0.1}$, with a typical brightness temperature of $10^4-10^5$ K. Its luminosity could be calculated accurately based on the observed emission line luminosities. The free-free emission may be best probed at $100-200$~GHz, before the thermal tail of the dust emission starts to dominate. 

If the radio is produced in magnetised plasma in the accretion disc corona, as suggested by the observed $L_{\rm R}/L_{\rm X}\simeq 10^{-5}$ relation, then a robust prediction is that a compact sub pc, optically thick, flat spectrum source should emerge at high radio frequencies. The mm emission is expected to originate from a few $R_g$, and thus overlap the X-ray emitting region. The emission in the two bands is expected to correlate perhaps according to the Neupert effect, where $L_{\rm R}=dL_{\rm x}/dt$, as detected in stellar coronae (see Sec.~\ref{sec:corona}). 

In analogy with coronally active stars, Coronal Mass Ejections (CMEs) may occur in the AGN coronae, which would also emit extended optically-thin radio emission in a form of outflowing blobs of highly magnetised plasma. This phenomenon would be more intense in very active coronae, which are expected to be in high-Eddington accretion discs (see \citealt{laor19}). If the CME speeds are highly subluminal, the jet model would coincide with the CME phenomenon and the jet base would physically coincide with the corona (see also \citealt{merloni02,liu14,king17}). 

Radio emission produced by the SF will inevitably form a smoothly distributed host-like extended emission, which is obviously non-variable, and characterised by a steep spectral slope. The SFR in AGN host galaxies in the local Universe may reach $\sim 300 M_{\odot}$~yr$^{-1}$. At such a high SFR the Supernovae rate reaches a few per year. This raises the exciting possibility of follow-up deep VLBI observations of nearby AGN ($z<0.1$), which show diffuse radio emission consistent with SF, and try to detect compact sources on the pc scale. 
It may be possible to detect individual SNe and SN remnants, as impressively done for the nearby ULIRGs M~82, Arp~220 and Arp~299-A \citep{muxlow94,batejat11,perez09}. If doable, it will allow us to measure directly the SFR rate of massive stars very close to the centre.

The cleanest diagnostic of the various possible emission processes is the comparison of X-ray and radio flux variations. Correlated and lagged variability will provide strong support for the radio and X-ray sources being physically related and specific radio/X-ray time lags might point to a common origin. If radio emission comes from a synchrotron jet, we expect an X-ray/radio correlation, such as that seen in XRBs (e.g. \citealt{corbel13}). Assuming a scaling with BH mass, a jet origin would predict the low frequency radio to lag behind the X-ray photons in RQ AGN by tens of days as found in the RQ NGC\,7213 (24 days radio lag, Fig~\ref{fig:bell11}, \citealt{bell11}). In contrast in the mm range, where the innermost disc becomes optically thin, one may observe the variability pattern seen in the solar corona, where magnetic field reconnection events lead to rapid particle acceleration and the production of radio emission, causing X-ray emission (cooling) related to the integral of the radio emission (heating) and to lag behind the radio. Conversely, in the case of SF, radio and X-ray emission will not vary on years time-scales. Winds occur at different spatial scales and on different physical conditions, detailed modeling for the expected radio/X-ray correlated variability is still at its dawn. 

\section{Near-future Desired Observations}

Significant improvement in this field could already be achieved with the current radio telescopes in a short-term future, ideally requiring: i) a complete and statistically meaningful sample of RQ AGN which covers a large range of radio luminosities and accretion rates, achievable by reaching $\mu$Jy flux-density levels with short exposure times (i.e., $\sim$ 30 min of total time with array-A VLA to reach a sensitivity of 10$\mu$Jy beam$^{-1}$ and 0.4$\arcsec$ resolution at 5 GHz; $\sim$ 1 hour for eMERLIN at 5 GHz for a sensitivity of 25$\mu$Jy beam$^{-1}$ and 0.04$\arcsec$ resolution); ii) a full spectral coverage from cm band to FIR to discriminate between the different emission mechanisms (i.e., combining cm-band VLA and mm-band ALMA data, see \citealt{doi11,doi13,inoue18}); iii) multi-scale (arcsec to mas-resolution) imaging to map and resolve the different emission structures from the galaxy to the innermost regions and trace their connection. In this respect, very-long baseline radio arrays have been implemented in the recent years with the possibility to combine them together in the Global VLBI network\footnote{The combination of the European VLBI (EVN)  and VLBA is known as Global VLBI (https://www3.mpifr-bonn.mpg.de/div/vlbi/globalmm/).}, allowing us to reach unprecedented angular resolution (up to 45 $\mu$arcsec) and {\it uv} coverage. An optimal technique, though not much exploited, is the combination of visibilities from short and long baselines obtained for example with VLA and eMERLIN. This procedure could probe intermediate scales in the {\it uv} plane, ideal for studying pc-scale extended radio emission in RQ AGN, which would be resolved out at long baselines and unresolved at short baselines.

The advent of the new radio facilities with unprecedented large collecting area, radio frequency coverage and angular resolution (i.e. LOFAR, SKA, ngVLA, EHT, ALMA, space VLBI; \citealt{vanhaarlem13,carilli15,murphy17,gurvitis18}), which will be $\sim$50 times more sensitive and 10$^{4}$ times faster in the case of the SKA (sensitivity of 0.4 mJy beam$^{-1}$ in one minute between 70 and 300 MHz), will represent the springboard to a new era of radio science. At $\mu$Jy-level flux densities, the fraction of RQ AGN significantly increases implying that large numbers of these
sources will be detected (a survey reaching $\sim 1~\mu$Jy at $\sim 1$ GHz over 30 deg$^2$ will detect 
$\approx 200,000$ RQ AGN: \citealt{Padovani_2016}). SKA1 will be able to map the sky and reach a brightness temperature of 0.2 K at 10 GHz with sensitivity of 15 $\mu$Jy arcsec$^{-2}$ for 1000 hr observations with a 0.1$\arcsec$ synthesised beam \citep{dewdeny13}. These values correspond to detecting galaxies forming stars at $\sim$100 and 10 M$_{\odot}$ yr$^{-1}$ at all redshifts and a radio luminosity of $\sim$10$^{33}$ erg s$^{-1}$ at z=0.3 \citep[1.6 Gpc;][]{murphy15}.
Future facilities will enable polarization studies of faint sources, that can reveal the geometry of the compact radio emission and favour or disfavour magnetically-driven mechanism. Long-baseline high sensitivity facilities will allow high positional accuracy measurements in order to detect the motions of sub-relativistic radio blobs (minimum detectable size corresponds to a length of $\sim$1.6 light-years  at z = 0.15 with a resolution of 1 mas, \citealt{paragi15})

Short time investment on the new radio facilities at high radio frequencies could resolve mas-scale $\mu$Jy-level radio components in galactic and extra-galactic sources. This will open up an uncharted radio sky, ready to be followed up by other new international facilities at shorter wavelengths (i.e., LSST, Euclid, WFIRST, ELTs, JWST, SPICA, eROSITA, Athena), providing multi-frequency monitoring campaigns with short and long visit cadences. 

\section{Summary and Outlook}

Radio astronomy is witnessing a golden age thanks to the large number of already available interferometers and to the high throughput of upcoming new facilities. RQ AGN become the dominant AGN radio population at flux densities below $\approx 0.1$ mJy at 1.4 GHz \citep[e.g.,][]{Padovani_2016}. However, it is not only about quantity, but also about quality: nuclei can be investigated in their exciting variety of physical mechanisms in act.  Accretion of material and consequent ejection of plasma, coronal activity, interaction with the surrounding medium and star-formation are the major sources of radio emission in this class of objects. Disentangling the above physical processes (or discovering new ones!) is difficult. Characteristic observational signatures can be identified for each process (although degeneracies are common) and testable predictions are within the reach of the current and future radio (including mm band) interferometers.
The effort that the different observational communities should undertake is to boost the multi-frequency approach, by simplifying the access to the time domain astronomy and to intensify the global observatory network to finally enable simultaneous multi-band campaigns. Furthermore, there is urgent need to bridge the different astrophysical communities perspectives, from stars to SMBHs: accretion, ejection, coronae, winds and star formation are basic ingredients in most accreting systems, as well as in the accretion history of the Universe.\\

\acknowledgments

This review is the result of several fruitful discussions raised during the meeting 'The radio--X-ray connection in accreting objects' (21-25 May 2018, Tenuta Monacelle, Monopoli, Italy). The authors wish to thank all the participants to the meeting: John Bally, Niel Brandt, Gianfranco Brunetti, Alessandro Capetti, Stephane Corbel, Lixin Dai, Jordy Davelaar, Barbara De Marco, Jonathan Ferreira, Jose Gomez, Martin Hardcastle, Yoshiyoki Inuoe, Miranda Jarvis, Preeti Kharb, Amy Kimball (in particular for her original idea that inspired Figure 2), Mike Koss, Carol Mundell, Dheeraj Pasham, Uria Peretz, Manel Perucho, Richard Plotkin, Isabella Prandoni, Juri Poutanen, Tim Roberts, David Williams, Clive Tadhunter, Sasha Tchekhovskoy, Francesco Ursini, Diana Worrall, and Nadia Zakamska. We acknowledge the useful comments by the referees that helped improving the manuscript. We also wish to thank Diego Altamirano, Piergiorgio Casella, Elia Chiaraluce, Sebastian Hoenig, Matt Middleton, Mayukh Pahari, Miguel P{\'e}rez-Torres and David Williams.

\appendix

\section{Acronyms}

\noindent
{\bf Atacama} Large Millimeter/submillimeter Arra\\
{\bf Athena} Advanced Telescope for High ENergy Astrophysics\\
{\bf EHT} Event Horizon Telescope\\
{\bf ELT} Extremely Large Telescope\\
{\bf EVN} European VLBI Network \\
{\bf eMERLIN} extended Multi-Element Radio Linked Interferometer Network\\
{\bf eROSITA} extended ROentgen Survey with an Imaging Telescope Array\\
{\bf FIRST} Faint Images of the Radio Sky\\
{\bf ISM} Interstellar medium\\
{\bf JWST} James Webb Space Telescope \\
{\bf LINER} Low-Ionization Nuclear Emission-line Region \\
{\bf LOFAR} LOw Frequency ARray\\
{\bf LSST} Large Synoptic Survey Telescope\\
{\bf NVSS} NRAO VLA Sky Survey\\
{\bf ngVLA} new-generation Very Large Array \\
{\bf QSO} Quasi Stellar Object \\
{\bf SKA} Square Kilometre Array\\
{\bf SPICA} Space Infrared Telescope for Cosmology and  Astrophysics\\
{\bf VLA} Very Large Array\\
{\bf VLBA} Very Long Baseline Array\\
{\bf VLBI} Very Long Baseline Interferometry\\
{\bf WFIRST} Wide-Field InfraRed Survey Telescope

\end{document}